\documentclass[12pt,final]{iopart}
\usepackage{amsfonts,amssymb,amsthm,array,amsmath,mathrsfs}
\usepackage{relsize}
\usepackage{cancel}
\usepackage{mathtools}
\usepackage[T1]{fontenc}
\usepackage{fancyhdr}
\usepackage{cite}

\usepackage{hyperref}
\usepackage{bibentry}
\usepackage[colorinlistoftodos]{todonotes} 

\usepackage{graphicx}% Include figure files
\usepackage{dcolumn}% Align table columns on decimal point
\usepackage{bm}% bold math
\usepackage{enumerate}
\usepackage{slashed}
\usepackage{simplewick}
\usepackage{comment}

\usepackage{soul}

\usepackage[english]{babel}
\usepackage{appendix}
\usepackage[utf8]{inputenc}

\usepackage{marginnote}
\usepackage[normalem]{ulem}

\usepackage{subfigure}
\newcommand{\sA}{\scriptscriptstyle\rm A}

\newcommand{\eq}[1]{(\ref{#1})}

\newcommand{\la}{\label}
\newcommand{\ba}{\begin{align}}
\newcommand{\ee}{\end{equation}}
\newcommand{\be}{\begin{equation}}

\def\12{\frac{1}{2}}
\newcommand{\p}{\partial}
\newcommand{\en}{\end{align}}
\newcommand{\ep}{\epsilon}

\usepackage{color}

\usepackage{amsmath}

\begin{document}

\title[Anomalies in fluid dynamics \ldots]{Anomalies in fluid dynamics: flows in a chiral background via variational principle 
}

\author{A.G.~Abanov}
%\affiliation
\address{Department of Physics and Astronomy and Simons Center for
Geometry
and Physics, Stony Brook University, Stony Brook, NY 11794, USA}

\author{P.B.~Wiegmann}
%\affiliation
\address{Kadanoff Center for Theoretical Physics, University of Chicago,
5640 South Ellis Ave, Chicago, IL 60637, USA
}

\date{\today}

\begin{abstract}
We study flows of barotropic perfect fluid under the simultaneous action of the electromagnetic field and the axial-vector potential, the external field conjugate to the fluid helicity. We obtain the deformation of the Euler equation by the axial-vector potential and the deformations of various currents by two external fields. We show that the divergence of the vector and axial currents are controlled by the \emph{chiral anomaly} known in quantum field theories with Dirac fermions. We obtain these results by extending the variational principle for barotropic flows of a perfect fluid by coupling with the external axial-vector potential. 
\end{abstract}

\maketitle\flushbottom

%%%%%%%%%%%%%%%%%%%%%%%%%%%%%%%%%%%%%
\section{Introduction}
%%%%%%%%%%%%%%%%%%%%%%%%%%%%%%%%%%%%%

In two recent papers, \cite{abanov2022axial,wiegmann2022chiral} we have shown that the axial-current (or chiral) anomaly broadly known in quantum field theories with Dirac fermions is also a kinematic property of barotropic flows of classical hydrodynamics.  

The axial-vector anomaly by Adler  \cite{adler1969axial}
and Bell and Jackiw \cite{bell1969pcac} states that parallel electric and magnetic field cause a divergence of the axial current. In units of the flux quantum
\(\Phi_0=he/c\), and the Planck constant \(h\), the divergence of the axial current reads \begin{align}
  \p\!\cdot\!j_{\sA}= 2 \bm E\!\cdot \!\bm B \,.
 \la{-1}
\end{align}
We recall that the axial current is the difference between the currents of chiral left and right components of the Dirac multiplet. Eq.\eq{-1} is referred to as \emph{anomaly} because it seemingly contradicts the axial gauge symmetry of the classical Dirac Lagrangian, which, by the Noether theorem, is expected to yield the divergence-free axial current. A resolution of this broadly studied puzzle is that Eq.\eq{-1}   reflects a property of the space of quantum states for which the group of axial transformations is not symmetry. For that reason, the axial-current anomaly is largely insensitive to interactions.

The hydrodynamic description of relativistic fluids consistent with chiral anomalies has been studied in the last decade see Refs.~\cite{son2009hydrodynamics,neiman2011relativistic,isachenkov2011chiral,nair2012fluids,stephanov2012chiral,son2012berry,banerjee2012constraints,jensen2013thermodynamics,haehl2014effective,dubovsky2014effective,monteiro2015hydrodynamics,mitra2022divergence} and references therein.

In Ref.~\cite{abanov2022axial} we showed that in fluids, the helicity current 
\begin{align}
        j_{\sA}^\alpha=\ep^{\alpha\beta\gamma\delta}p_\beta\p_\gamma p_\delta
\end{align} 
plays the role of an axial current. In this formula, \(p_\alpha=(p_0,\bm p)\)
is the fluid's 4-momentum per particle. If the fluid is relativistic, this is the usual 4-momentum. If the fluid is Galilean invariant,  then \(\bm p=m\bm v\) is the fluid momentum, and $p_0$ is the energy per particle 
\begin{align}
 \la{3.1}
        p_0=-\p\mathcal{E}/\p n\,, \qquad \mathcal{E}=n\bm p^2/(2m)+\varepsilon(n)\,.
\end{align}
Here \(n\) is the particle number, and $\varepsilon(n)$ is the internal energy of the fluid per unit volume. In a perfect barotropic fluid, the helicity current is conserved \(\p\!\cdot\! j_{\sA}=0\) \cite{carter1979perfect}. However, if the fluid is charged and is placed in electromagnetic field, a deformed current of fluid chirality 
\begin{align}
 \la{2}
        j_{\sA} ^\alpha=\ep^{\alpha\beta\gamma\delta}p_\beta(\p_\gamma
p_\delta+F_{\gamma\delta})\,
\end{align}   
obeys the anomaly equation \eq{-1}. In this formula, \(F_{\alpha\beta}=\p_\alpha A_\beta\!-\!\p_\beta A_\alpha\) is the electromagnetic field tensor, and \(A_\alpha \) is the electromagnetic vector potential.  

Similarly to fermions, one may introduce an external axial potential \(A^{\sA}=(A^{\sA}_0,\bm A^{\sA})\) conjugate to the  chirality current \eq{2}  by adding the term \(A^{\sA}\cdot\! j_{\sA}\) to the fluid action. Such potential produces a flow with non-zero fluid chirality which has interesting properties addressed in \cite{wiegmann2022chiral}.

A related problem for fermions was a subject of renewed interest (see Ref.~\cite{arouca2022quantum} for a recent review). It had been found that the axial potential can not be applied to an isolated system. The system must be connected to a reservoir which supplies an electric charge (but not fluid particles) so that the electric current carried by fluid particles diverges as 
\begin{align}
  	\p\!\cdot\! j = 2(\bm E^{\sA}\!\cdot \!\bm B+\bm B^{\sA}\!\cdot \!\bm E)\,,
 \la{-2}
\end{align}  
where \(\bm E^{\sA}=\bm \nabla A_0^{\sA}\!-\!\dot{\bm A}^{\sA}\) and \(\bm B^{\sA}=\bm\nabla\!\times \!\bm A^{\sA}\) are the axial analogs of electric and magnetic fields.

Here we show that the same formula \eq{-2} also holds in fluids.
In a particular case, when the axial potential has only
a temporal component \(A_\alpha^{\sA}=(A_0^{\sA},0)\),  with $A_0^{\sA}$ playing a role of the  axial chemical potential, had been considered in Ref.~\cite{wiegmann2022chiral}.  

A barotropic fluid with or without axial coupling belongs to a special class of hydrodynamic systems referred to as \emph{uniformly canonical} \cite{carter1979perfect,carter1988standard}. Flows of such systems possess a special geometric property: the conservation of the helicity current (for a review of conservation laws in uniformly canonical systems, see \cite{markakis2017conservation}). 

In this paper, we extend the results of  \cite{wiegmann2022chiral} to include all components of the axial potential. For this purpose,  a spacetime-covariant variational principle for hydrodynamics is the most convenient. We introduce the axial coupling by extending the variational principle known for a perfect fluid. See, also Ref.~\cite{monteiro2015hydrodynamics} for a use of the variational functional for a related problem.

We focus on the derivation of equations of motions, and the hydrodynamic representation of vector and axial currents, leaving an important question of the physical realization of the axial coupling in fluids open.

%%%%%%%%%%%%%%%%%%%%%%%%%%%%%%%%%%%%%
\section{Coupling with the axial potential}
 \label{sec:vp1}
%%%%%%%%%%%%%%%%%%%%%%%%%%%%%%%%%%%%%

We will introduce the axial coupling via the deformation of covariant variational functional for a barotropic fluid. However, before this, let us take a moment to review the variational approach for a broader class of uniformly canonical fluid systems, of which the barotropic fluid is the simplest example.

Uniformly canonical systems whose significance was emphasized by Carter \cite{carter1979perfect,carter1988standard} (see, \cite{markakis2017conservation} for a review) could be defined as a system whose variational functional depends on no other dynamical variables but 4-momentum $p_\alpha$, or the canonical 4-momentum
 \begin{align}
    \pi_\alpha=p_\alpha+A_\alpha\,
\end{align}  
 if the system is charged and is placed in an electromagnetic field. This requirement, in particular, excludes baroclinic fluids where the rest energy \(\varepsilon\) in \eq{3.1} also depends on another Lagrangian scalar, such as the entropy (see, the Sec.\ref{D}).

 A variational functional for a perfect fluid  is known for a long time \cite{seliger1968variational,schutz1970perfect}. It is given by the spacetime integral of the fluid pressure
\begin{align}
 \la{6.1}
    \mathcal{A}^{(0)}=\int P(p)\, d^4 x\,.
\end{align}  
In this approach, pressure, being a function of the chemical potential $P(\mu)$, is expressed through the relativistic 4-momentum according to the kinematic relation \(\mu^2 = -p^2\), where \(\mu\) is the
chemical potential. In the non-relativistic fluid, this relation is given by Eq.\eq{3.1} so that $\mu = -p_{0}-\bm{p}^{2}/2m$. In the electromagnetic field, the action (\ref{6.1}) retains its form but should be considered as a functional of the canonical fluid 4-momentum $\pi_\alpha$.  

We introduce the axial coupling by deforming the variational functional \eq{6}  
\begin{align}
 \la{7.1}
    \mathcal{A=A}^{(0)}+\int (A^{\sA}\cdot\! j_{\sA} )\,d^4x\,.
\end{align}
While the integrand of the first term in \eq{7.1} depends only on the momentum, the coupling also depends on momentum derivatives. As a result, the coupling significantly changes the relationship between the momentum, particle number current, and electric current. We discuss this in the next section. 

%%%%%%%%%%%%%%%%%%%%
\section{Currents}
 \la{C}
%%%%%%%%%%%%%%%%%%%%

The variational functional generates various currents which play an important role. The first is the particle number current defined as the conjugate to the canonical momentum. 
\begin{align}   
        \mathcal{J}^\alpha := -{\delta\mathcal{A}}/{\delta \pi_\alpha} \,.
 \label{Jdef100}
\end{align}
The particle number current satisfies the continuity equation (see Sec.~\ref{subsec-geom})
\begin{align}
        \p_\alpha\mathcal{J}^\alpha =0\,.
 \label{varphi100}
\end{align}

Next is the electric current defined by
\begin{align}
        J^\alpha:= {\delta\mathcal{A}}/{\delta A_\alpha}\,\Big|_\pi.
 \la{12.1}
\end{align} 
We emphasize that the variation is taken at a fixed \(\pi\). As required by the gauge invariance,
 \be 
        A_\alpha\to A_\alpha+\p_\alpha\varphi,\quad \pi_\alpha\to \pi_\alpha+\p_\alpha\varphi \,,
 \la{10}
\ee 
the electric current, is also, divergence free
\begin{align}
        \p_\alpha {J}^\alpha =0\,.
 \la{13}
\end{align}
And, finally, the axial current 
\begin{align}
        j_{\sA}^\alpha := \delta\mathcal{A}/\delta A_\alpha^{\sA}\,.
 \la{14.1}
\end{align}
If there were no axial potential, the particle number current \(\mathcal{J}\) and the electric current \(J\) would be equal and equal to \((n,\, n\bm p/m)\). The axial coupling changes this accidental relation.

%%%%%%%%%%%%%%%%%%%%
\section{Variational principle  for uniformly canonical fluid  system}
%%%%%%%%%%%%%%%%%%%%

The  variational principle states that on equations of motions  
\begin{align}
        \delta \mathcal{A}=0
 \la{7}
\end{align}
under admissible variations of the 4-momentum $\pi$.

Since we assume that  there are no other variational parameters but \(\pi\), the variational principle for the uniformly canonical system requires
\begin{align}
 \la{6}
        -\int \mathcal{J}^\alpha \delta\pi _\alpha\, d^4x=0\,.
\end{align}

It is  known that in hydrodynamics one can not vary the momentum freely. 
 The canonical 4-momentum is a spacetime covector. The admissible variations of the momentum are those which could be obtained  by  spacetime  diffeomorphisms \(x\to x+\ep(x)\). Under spacetime diffeomorphisms, a covector transforms as (see, e.g., \cite{carter1979perfect,volovik1979poisson,rieutord2006introduction}) 
\begin{align}
    \delta \pi_\alpha &= \ep^\beta\p_\beta \pi_\alpha +\pi_\beta\p_\alpha \ep^\beta 
 \nonumber \\
    &= \ep^\beta\Omega_{\beta\alpha}+\p_\alpha(\ep^\beta\pi_\beta)\,, 
 \label{admvar}
\end{align} 
where we introduced the 4-canonical vorticity tensor as
\be 
        \Omega_{\alpha\beta}=\p_\alpha\pi_\beta-\p_\beta\pi_\alpha \,.
\ee  
The variations of the canonical momentum (\ref{admvar}) for an arbitrary infinitesimal vector field $\ep^\beta$ are the admissible variations.

%%%%%%%%%%%%%%%%%%%%%%%%%%%%%%%%%%%%%
\subsection{Carter-Lichnerowicz equation for uniformly canonical systems}
\label{subsec-geom}

Substituting (\ref{admvar}) into (\ref{6}) and integrating by parts, we derive the relation
\begin{align}
    	&\mathcal{J}^\alpha\Omega_{\alpha\beta} +\pi_\beta\p_\alpha \mathcal{J}^\alpha =0\,.
 \label{vare101}
\end{align}    
Contracting this equation with $\mathcal{J}^\beta$ cancels the first term. Then the continuity equation \eq{varphi100}| follows under the assumption $\mathcal{J}^\beta\pi_\beta\neq 0$. Then, taking into account the continuity equation \eq{varphi100}, we obtain form \eq{vare101} the equations of motion in the Carter-Lichnerowicz form  
\begin{align}
    \mathcal{J}^\alpha \Omega_{\alpha\beta}  =0\,.
 \label{CL100}
\end{align}
We remark that for uniformly canonical systems the continuity equation (\ref{13}) follows from the variational principle (\ref{6},\ref{admvar}). This occurs because a particular diffeomorphism along the direction of the particle current $\epsilon^\beta = \varphi \mathcal{J}^\beta/(\mathcal{J}\!\cdot\pi)$, where $\varphi(x)$ is an infinitesimal spacetime scalar, acts as a gauge transformation. Under such variation, (\ref{6}) reads $\mathcal{\delta \mathcal{A} =-\int J}^\alpha \p_\alpha\varphi \,d^4x=\mathcal{\int \varphi\,  \p_\alpha J}^\alpha d^4x$. Then the variation over $\varphi$ gives the continuity equation (\ref{varphi100}). This does not happen, for example, in baroclinic fluid.

We also remark that the Carter-Lichnerowicz equation (\ref{CL100}) does not depend on a spacetime structure. It is valid for relativistic and non-relativistic fluids alike. The spacetime metric enters as a relation between \(\mathcal{J}\) and \(\pi\) encoded by the form of the variational functional $\mathcal{A}$.

%%%%%%%%%%%%%%%%%%%%%%
\section{Conservation of helicity current and axial current anomaly}
%%%%%%%%%%%%%%%%%%%%%%

%%%%%%%%%%%%%%%%%%%%%%
\subsection{Conservation of helicity current } \la{5.1}

An essential property of uniformly canonical systems is the conservation of the helicity current 
\cite{carter1979perfect,khesin1989invariants,bekenstein1987helicity,arnold2008topological}. Helicity current is defined as a vector  dual to the 3-form $\pi \wedge d\pi$
\be 
        h^\alpha =\ep^{\alpha\beta\delta\gamma}\pi_\beta\p_\delta \pi_\gamma\,.
 \la{25}
\ee
Helicity is conserved 
\begin{align}
        \p_\alpha h^\alpha=0
 \la{27}
\end{align}
as a consequence of \eq{CL100} independently of the relation between the current \(\mathcal{J}\) and the momentum \(\pi\) and with or without electromagnetic and axial fields. 

We briefly derive this important fact. It is convenient to work with differential forms. We denote the canonical momentum 1-form as $\pi=\pi_\alpha dx^\alpha$ and the closed  2-form of canonical vorticity as \(\Omega=\frac{1}{2}\Omega_{\alpha\beta}dx^\alpha\wedge d x^\beta=d\pi \). Then, the helicity 3-form reads  \(h = \pi\wedge d\pi\).  The divergence of the helicity current (\ref{25}) is dual to the 4-form $dh = d\pi\wedge d\pi =\Omega\wedge \Omega$. 
    
Let us show that on a solution of the equation \eq{CL100} the helicity form is closed \(dh=0\).  Eq.~\eq{CL100} states that vorticity surfaces defined by $\Omega$ are normal to the
current. Hence, the vorticity 2-form cannot have a maximal rank.  At the same time, the 4-form $\Omega\wedge \Omega$ is a top form in four dimensions and should be proportional to the volume form. As it does not have a maximal rank four, the proportionality coefficient must be zero. Then
\begin{align}
        dh=\Omega\wedge\Omega=0 \,.
 \la{OO}
\end{align}
In tensor notations, one uses the identity 
\begin{align}\la{24}
        2\Omega_{\beta\alpha}\,{}^\star \Omega^{\alpha\lambda}
        =\tfrac 12\delta_\beta^ \lambda \Omega_{\gamma\alpha} \,{}^\star \Omega^{\alpha\gamma}\,,
\end{align}
valid for any antisymmetric 2-tensor, where we denoted  \(^\star
\Omega^{\alpha\beta}=\tfrac 12 \ep^{\alpha\beta\gamma\delta}\Omega_{\gamma\delta}\). Multiplying the identity by $\mathcal{J}^\beta$ and using \eq{CL100}, one arrives at $\Omega_{\gamma\alpha} \,{}^\star \Omega^{\alpha\gamma}=0$, which is equivalent to \eq{OO}.

Alternatively, one can derive (\ref{27}) by choosing a special diffeomorphism in the variational principle (cf., \cite{webb2014local}). In (\ref{6},\ref{admvar}), we set $\epsilon^\alpha = 2\eta h^\alpha/(\mathcal{J}\!\cdot \pi)$, where $h^\alpha$ is the helicity current and  $\eta(x)$ is an infinitesimal spacetime pseudo-scalar.  Then using identity \eq{24}, we obtain $\delta\mathcal{A} = -\int \eta\, dh\,$. Setting it to zero, we obtain \eq{27}.

%%%%%%%%%%%%%%%%%%%%%%
\subsection{Axial current anomaly}

Helicity current \eq{25}, although conserved, is not a gauge invariant entity. Under a gauge transformation \eq{10} $h^\alpha\to h^\alpha+\p_\beta({}^\star\Omega_{\alpha\beta} \varphi)$. Counter to the total helicity \(\mathcal{H}=\int h^0 d^3x\), helicity current can not be associated with an Eulerian observable. This obstacle is the signature of an anomaly. It  turns out that a modification of the helicity current in such a way that it remains divergence-free and at the same time gauge invariant is not possible.  A deformation of the helicity current to the chirality current
\begin{align}
    j_{\sA}^\alpha = \epsilon^{\alpha\beta\gamma\delta} 
    p_\beta (\p_\gamma p_\delta +F_{\gamma\delta})\,
 \label{jAdef200}
\end{align}
solves the problem at the expense of its conservation. However, as we showed in  \cite{abanov2022axial}, the divergence of the chirality current  does not involve dynamical fields. It is expressed solely in terms of the electromagnetic field as in \eq{-1}. 

Differential forms shorten calculations. We represent the chirality current \eq{jAdef200}
as a dual of a 3-form 
\be\la{26}
j_{\sA}=p\wedge d(p+2A) 
\ee 
and express it through the canonical momentum 1-form
$$j_{\sA}=(\pi-A)\wedge d(\pi+A)=h -A\wedge dA +d(A\wedge\pi)\,.$$ The divergence of the chirality current is dual to the exterior derivative of the chirality 3-form. The exterior derivative eliminates the last term in the above expression, which is the exact form. Hence, we have
\begin{align}
    	&dj_{\sA}=dh-dA\wedge dA\,.
\end{align}
Then the helicity conservation (\ref{27}) yields the axial anomaly $dj_{\sA}=-dA\wedge dA$ 
quoted  in the introduction as \eq{-1}
\begin{align}
    	\p_\alpha j_{\sA}^\alpha 
    	= \frac{1}{2}F_{\alpha\beta} {}^\star\!F^{\beta\alpha}\,,
 \label{divjA200}
\end{align}
where  $\,{}^\star\! F^{\alpha\beta}=\frac{1}{2} \epsilon^{\alpha\beta\gamma\delta} F_{\gamma\delta}$ is the dual electromagnetic field tensor.

In the remaining part of the paper, we refer to chirality current $j_{\sA}$ as axial current.

%%%%%%%%%%%%%%%%%%%%%
\section{Equations of motion in the form of conservation laws}
%%%%%%%%%%%%%%%%%%%%%

Now we obtain the equations of motions in the form of the conservation law generated by spacetime diffeomorphisms. These equations are equivalent to the Carter-Lichnerowich equations and the continuity equation but emphasize different aspects of flows in the chiral background. We start from the conventional barotropic fluid.

%%%%%%%%%%%%%%%%%%%%%
\subsection{Perfect barotropic fluid}
 \la{PF}

In this section, we consider a perfect barotropic fluid not coupled to external potentials (neither vector nor axial field). In this case, the canonical momentum and the fluid momentum are identical \(\pi=p\) as \(A=0\). In the absence of axial field, the particle number current is the partial derivative of pressure with respect to the momentum $-\frac{\p P}{\p p_\alpha}$.  It is equal to 
\begin{align}
    -\frac{\p P}{\p p_\alpha}=nu^\alpha\,,
 \label{Jdef105}
\end{align}
where we denote \(nu^\alpha=(n,\, n\bm v)\) and \(m \bm v= \bm p \). 

To obtain \eq{Jdef105}, we use the thermodynamic relation \(\delta P=n\delta \mu\) for the barotropic fluid, where
 \(\mu=d\varepsilon/d n\) is the chemical potential of the fluid. Then, the variational form of the relation \eq{3.1}  reads \(-\delta \mu=\delta \pi_0+(\bm p/m)\delta \bm p=u^\alpha \delta p_\alpha\). It gives  \eq{Jdef105}. Having a relation between the particle number current and momentum \eq{Jdef105}, the Carter-Lichnerowicz equation \eq{CL100} reads 
\begin{align}
        \dot{\bm p}-\bm \nabla p_0-\bm v\times (\bm\nabla\times\bm p)=0\,.
\end{align}
This is, of course, known as the Euler equation.

We can cast this equation in the form of a conservation law. With the help of (\ref{admvar}) with $\pi=p$, we compute  the variation of  $\delta\mathcal{A}^{(0)}$ as 
\begin{align}
	\delta\mathcal{A}^{(0)}=\int  \frac{\p P}{\p p_\alpha} \delta p_\alpha d^4x
	=\int\left[\frac{\p P}{\p p_\alpha}\p_\beta p_\alpha+\p_\alpha
	\left(-\frac{\p P}{\p p_\alpha}p_\beta\right )\right ]\ep^\beta d^4 x\,.
\end{align}
The first term in this formula is the complete derivative  $\frac{\p P}{\p p_\alpha}\p_\beta p _\alpha=\p_\beta P$. 
Then
\begin{align}
 \la{29} 
        \delta\mathcal{A}^{(0)}=\int \p_\alpha T^\alpha_\beta\, \ep^\beta\, d^4 x\,,
\end{align}
where 
\begin{align}
        T^\alpha_\beta &= {nu}^\alpha p_\beta +P\delta^\alpha_\beta\,
 \la{T999}
\end{align}
is the momentum-stress-energy tensor of the perfect fluid. We recognize the density of the variational functional  $P$ as fluid pressure. Equating the variation \eq{29} to zero, we obtain the conservation law equations  
\begin{align}
    \p_\alpha T^\alpha_\beta = 0.
 \label{T1000}
\end{align}
The formulas (\ref{Jdef105},\ref{29},\ref{T999},\ref{T1000}) also hold for relativistic fluid.

%%%%%%%%%%%%%%%%%%%%%
\subsection{Barotropic fluid with the axial-vector potential}

Now we extend the equations of motions \eq{T1000} by coupling with the axial-vector and electromagnetic potentials. We will use the general equations 
\begin{align}
 \la{32}
        \p_\alpha T^\alpha_\beta = F_{\beta\alpha}J^\alpha- A_\beta \p_\alpha J^\alpha+F^{\sA}_{\beta\alpha}j_A^\alpha- A_\beta ^{\sA}\p_\alpha j_{\sA}^\alpha\,
\end{align}
and the divergence of the currents obtained above. 

The electric current is, of course, divergence-free \eq{13}, so the second term is dropped from the r.h.s. of \eq{32}, but the divergence of the axial current is given by  the axial current anomaly \eq{divjA200}. With the help of the identity 
\begin{align}
        2F_{\beta\alpha}\,{}^\star\! F^{\alpha\lambda}=\tfrac 12\delta_\beta^ \lambda 
        F_{\gamma\alpha} \,{}^\star\! F^{\alpha\gamma} \,,
\end{align} 
we rewrite the last term in the r.h.s. of \eq{32} as \(A_\beta ^{\sA}\p_\alpha j_{\sA}^\alpha=-2F_{\beta\alpha}
\,{}^\star\!
F^{\alpha\gamma}A_\gamma ^{\sA}\) and combine it with the first term. We obtain one of the main results of this paper 
\begin{align}
 \la{39}
        \p_\alpha T^\alpha_\beta = F_{\beta\alpha}j^\alpha+F^{\sA}_{\beta\alpha}j_A^\alpha\,,
\end{align} 
where we introduced the \emph{vector current}
\begin{align}
 \la{40}
         j^\alpha=J^\alpha-2A_\gamma^{\sA} {}^\star\!F^{\alpha\gamma}\,.
\end{align} 

To complete this section, we comment on the derivation of Eq.\eq{32}.  Since \(j_A \) is a dual of the 3-form \eq{26}, it is an (axial) vector-current. Then, the coupling \(\int A^{\sA}\cdot j_{\sA}\, d^4x\) does not change under spacetime diffeomorphisms  \eq{admvar} and  the simultaneous transformation of the axial potential treated  as a spacetime covector  
\begin{align}
 \la{36}
        \delta A_\alpha ^{\sA} =  \ep^\beta\p_\beta A^{\sA}_\alpha +A^{\sA}_\beta \p_\alpha \ep^\beta\,.
\end{align}
Similarly,  if the vector  potential is transformed as a covector  
\begin{align}
 \la{34}
        \delta A_\alpha &=  \ep^\beta\p_\beta A_\alpha +A_\beta \p_\alpha \ep^\beta\,,
\end{align} 
the fluid momentum \(p_\alpha=\pi_\alpha-A_\alpha\) also transforms as a covector 
\begin{align}
 \la{35}
        \delta p_\alpha &=  \ep^\beta\p_\beta p_\alpha +p_\beta \p_\alpha \ep^\beta \,.
\end{align}
Let us now apply the combined transformations (\ref{admvar},\ref{36},\ref{34}) to the variational functional. We obtain
\begin{align}
    \delta \mathcal{A} = \int \, \left(-\mathcal{J}^\alpha\delta \pi_\alpha 
    +J^\alpha\delta A_\alpha + j_{\sA}^\alpha\delta A^{\sA}_\alpha \right) d^4x\,,
 \label{366}
\end{align}
where we used the definitions of the currents (\ref{Jdef100},\ref{12.1},\ref{14.1}). Then we evaluate \eq{366} on the equations of motions. This amounts to dropping the first term in the r.h.s. of \eq{366}. The remaining two terms we write with the help of (\ref{36},\ref{34}) as
 \begin{align} 
        &\delta \mathcal{A} = \int e^\beta \left(F_{\beta\alpha}J^\alpha-A_\beta\p_\alpha J^\alpha 
        +F^{\sA}_{\beta\alpha}j_{\sA}^\alpha-A^{\sA}_\beta\p_\alpha j_{\sA}^\alpha  \right)d^4x \,.
 \la{37}
\end{align}
At the same time, under combined transformations \(\delta \mathcal{A} =\delta \mathcal{A} ^{(0)}\), because the variation of the coupling term in \eq{7.1} drops. But \(\delta \mathcal{A} ^{(0)}\)  varies solely under \eq{35} as its integrand is a function of $p$. That variation was already computed in the previous section \eq{29}. Equating \eq{29} and \eq{37}, we obtain the main equation \eq{32}.

%%%%%%%%%%%%%%%%%%%
\section{Hydrodynamic representations of currents}
%%%%%%%%%%%%%%%%%%%

We obtained Eq.\eq{39}  before specifying the explicit form of the currents entering this equation. Now with the help of the explicit form of the variational functional, we compute the currents in terms of Eulerian fields. This gives us a hydrodynamic representation of the currents and closes the equations of motion given by two equivalent forms \eq{39} and \eq{CL100}. 

Using the general definitions of Sec.~\ref{C}, we obtain the deformation of the particle number current, electric current, and the vector current \eq{40}   
\begin{align}
        & \mathcal{J}^\alpha = nu^\alpha - {}^\star\!F^{\alpha\beta}_{\sA} p_\beta
        +2\,{}^\star \Omega^{\alpha\beta}A^{\sA}_\beta \,,
 \label{43} \\
        &J^\alpha =n u^\alpha  +{}^\star \! F_{\sA}^{\alpha\beta} p_\beta
        +2{}^\star \! F^{\alpha\beta}A^{\sA}_\beta
        =\mathcal{J}^\alpha  +2 \epsilon^{\alpha\beta\gamma\delta} 
        \p_\beta (A^{\sA}_\gamma  p_\delta) \,, 
 \\
        &j^\alpha=n u^\alpha  +{}^\star \! F_{\sA}^{\alpha\beta}p_\beta =J^\alpha  -2{}^\star \! F^{\alpha\beta}A^{\sA}_\beta\,.
 \label{44}
\end{align}
We notice  that particle number current is no longer equal to electric current, and both are different from \(nu^\alpha\), the current of the perfect fluid. The currents \(\mathcal{J}\) and  \(J\) are both divergence-free being different by a curl. However, the vector current \eq{44} diverges. We collect the divergence of all four currents here
\begin{align}
    &\p_\alpha \mathcal{J}^\alpha = 0,\quad \p_\alpha {J}^\alpha = 0\,,
 \label{45} \\
    &\p_\alpha j_{\sA}^\alpha = \tfrac{1}{2}F_{\alpha\beta} {}^\star\! F^{\beta\alpha}\,,
 \la{46}  \\
	&\p_\alpha j^\alpha = F_{\alpha\beta} {}^\star\!  F^{\beta\alpha}_{\sA} \,.
 \la{47}
\end{align}
Eq.\eq{47} is the chiral anomaly \eq{-2} quoted in the introduction. The relation between the vector current and the electric current \eq{44} suggests that the fluid is coupled with a charge reservoir which supplies and removes electric charge as required by the axial potential. The current contributed by the reservoir is $2\,{}^\star \! F^{\alpha\beta}A^{\sA}_\beta$. It does not depend on the fluid momentum. Together with the vector current $j$, it forms the electric current $J$. Hence, the vector current should be interpreted as the part of the electric current carried by fluid particles. The total electric current is conserved, but the parts are not. This interpretation was suggested in \cite{wiegmann2022chiral}. A few additional comments are in order. 

The  four equations (\ref{39}) and two equations for the divergence of the vector and the axial current (\ref{46},\ref{47}) determine four independent fields. They are  $\pi_\alpha$ or the density $n$ together with the 3-fluid momentum \(\bm p\). The compatibility of this formally over-constrained system is noteworthy. 

Having the explicit form of the currents, one can check (with a good deal of algebra) that equations of motions in the form of energy-momentum conservation laws \eq{39} are identical to the Carter-Lichnerowicz equation \eq{CL100} combined with the continuity equation \eq{varphi100}.  Eq.\eq{39} is explicitly invariant under the axial gauge transformation \(A^{\sA}_\alpha\to A^{\sA}_\alpha+\p_\alpha\varphi^{\sA}\), as the equation contains only Eulerian fields, the axial field tensor, and the vector and the axial current. In turn, these currents depend solely on Eulerian fields and the field tensors. It is less evident that equation \eq{CL100} also possesses this property since the particle number current explicitly depends on $A^{\sA}$ through the last term in \eq{43}. However, this term drops from equation \eq{CL100} due to the identity \eq{24} and the helicity current conservation \eq{OO}.
 
In the literature on relativistic hydrodynamics, it is customary to use  the following notations for relativistic vorticity and the projection of the electromagnetic field onto the hyperspace normal to  $u^\alpha$
\begin{align}
    	\omega^\alpha = \tfrac{1}{2}\epsilon^{\alpha\beta\gamma\delta}u_\beta\p_\gamma u_\delta,
	\quad  \mathcal{\mathrm{B}}^\alpha 
	= \tfrac{1}{2}\epsilon^{\alpha\beta\gamma\delta}u_\beta F_{\gamma\delta}\,.
\end{align}
In these notations, the vector and the axial currents read (compare to \cite{son2009hydrodynamics,monteiro2015hydrodynamics})
\begin{align}
    j^\alpha &= nu^\alpha + \mu \mathrm{B}_{\sA}^\alpha , \quad 
    j_{\sA}^\alpha = 2\mu^2 \omega^\alpha +2\mu\mathrm{B}^\alpha\,.
 \label{jAdef400}
\end{align}

Formulas (\ref{39},\ref{43}-\ref{47}) were announced in Ref.~\cite{wiegmann2022chiral}. Their derivation from the spacetime-covariant variational principle is the main result of this work.

%%%%%%%%%%%%%%%%%%%%%%%%%%%%%%%%%%%%%
\section{Discussion}
\label{D}
%%%%%%%%%%%%%%%%%%%%%%%%%%%%%%%%%%%%%

The chiral anomaly we discuss in this paper and in \cite{{wiegmann2022chiral}} occurs in barotropic fluids, or more generally, in uniformly canonical fluid systems. For example, in baroclinic fluids, there is no notion of the axial current since, in this case, the helicity current does not conserve. In a baroclinic fluid, the pressure depends on an additional thermodynamic variable, a Lagrangian invariant, such as entropy (per particle) \(S\). Hence, the variational functional for the perfect fluid
 \( \mathcal{A}^{(0)} = \int \, P(p,S)\,d^4x\) \cite{schutz1970perfect},
and, therefore, the entire functional now depends on five Eulerian fields $\pi_\alpha$ and $S$ with admissible variations (\ref{admvar}) supplemented by
\begin{align}
 	\delta S = \ep^\alpha \p_\alpha S\,.
 \la{admvar2}
\end{align}
Equation (\ref{vare101}) is then modified as
\begin{align}
 	\mathcal{J}^\alpha\Omega_{\alpha\beta} +\pi_{\beta}\p_{\alpha}\mathcal{J}^{\alpha}
	= nT\p_\beta S\,.
 \label{CL395}
\end{align}
Now multiplying by $\mathcal{J}^{\beta}$, we obtain 
\begin{align}
	(\pi\!\cdot \!\mathcal{J}) \p_{\alpha}\mathcal{J}^{\alpha} = nT (\mathcal{J}\!\cdot \!\p S)\,.
 \la{CL396}
\end{align}
Here $T=\p\varepsilon/\p S$, the temperature, is a thermodynamic variable conjugate to \(S\). It is important to notice here that the continuity equation (\ref{varphi100}) does not follow from the admissible variations \eq{admvar} as it was in the barotropic case discussed in Sec.~\ref{subsec-geom}. The continuity equation must be imposed outside of the variational principle. If one wants, however, to obtain the continuity equation within a variational principle approach, one has to extend the set of admissible variations (\ref{admvar},\ref{admvar2}) by the gauge transformation
\begin{align}
	\delta_\varphi\pi_{\alpha} =\p_{\alpha}\varphi\,, \qquad \delta_\varphi S =0\,.
 \la{admvar3}
\end{align}
This variation with an arbitrary $\varphi(x)$ ensures the continuity equation (\ref{varphi100}). Then (\ref{CL396}) and (\ref{CL395}) yield the transport equation for the entropy
$\mathcal{J}^{\alpha}\p_{\alpha}S =0$ and  the Carter-Lichnerowicz equation for baroclinic fluid \cite{markakis2017conservation}
\begin{align}
 	\mathcal{J}^\alpha\Omega_{\alpha\beta} = nT\p_\beta S\,.
 \label{CL400}
\end{align}
This equation is no longer uniformly canonical, that is, the r.h.s. of (\ref{CL400}) is not zero, and we cannot use the arguments of Sec.~\ref{5.1}. Instead, we derive
\begin{align}
	\p_\alpha h^\alpha
	= \frac{2 n T}{(\mathcal{J}\!\cdot\!\mathcal{J})}\mathcal{J}_\alpha \,
	{}^\star\Omega^{\alpha\beta}\p_\beta S \,.
\end{align}
We see that unless the flow is homentropic (entropy is constant in space and time), helicity current does not conserve. Hence, the notion of the axial field can not be extended beyond uniformly canonical fluid systems.

To summarize, in this work, we showed that the variational principle greatly facilitates the study of flows in a chiral background and studies of chiral anomalies in hydrodynamics. We obtained various forms of the equations of motions in the background potentials and showed that the vector and axial currents generated by the background potentials do not conserve. Rather they obey equations (\ref{-1},\ref{-2}) analogous to a chiral quantum anomaly for Dirac fermions.

%%%%%%%%%%%%%%%%%%%%%%%%%%%%%%%%%%%%%
\ack
%%%%%%%%%%%%%%%%%%%%%%%%%%%%%%%%%%%%%

The authors thank V.~P. Nair and G.~M. Monteiro for numerous discussions. The work of P.~W. was supported by the NSF under Grant NSF DMR-1949963. The work of A.~G.~A. was supported by the NSF under Grant NSF DMR-2116767.
The authors are grateful for the hospitality and support from Galileo Galilei Institute during the program on ``Randomness, Integrability, and Universality'' where this work was completed.
P.~W. also acknowledges the support of the Simons Foundation program for GGI Visiting Scientists. P.~W. also thanks the International Institute of Physics in Natal  Brazil for the hospitality.
 %%%%%%%%%%%%%%%%%%%%%%

%\newpage
%%%%%%%%%%%%%%%%%%%%%%%%%%%%%%%%%
%%%%%%%%%%%%%%%%%%%%%%%%%%%%%%%%%
\section*{References}
%%%%%%%%%%%%%%%%%%%%%%%

%\bibliographystyle{unsrt}
\bibliographystyle{iopart-num}

\bibliography{helicity}

\providecommand{\newblock}{}
\begin{thebibliography}{10}
\expandafter\ifx\csname url\endcsname\relax
  \def\url#1{{\tt #1}}\fi
\expandafter\ifx\csname urlprefix\endcsname\relax\def\urlprefix{URL }\fi
\providecommand{\eprint}[2][]{\url{#2}}

\bibitem{abanov2022axial}
Abanov A~G and Wiegmann P~B 2022 {\em Phys. Rev. Lett.\/} {\bf 128} 054501

\bibitem{wiegmann2022chiral}
Wiegmann P~B and Abanov A~G 2022 {\em JHEP\/} {\bf 06} 038

\bibitem{adler1969axial}
Adler S~L 1969 {\em Phys. Rev.\/} {\bf 177} 2426

\bibitem{bell1969pcac}
Bell J~S and Jackiw R 1969 {\em Il Nuovo Cimento A (1965-1970)\/} {\bf 60}
  47--61

\bibitem{son2009hydrodynamics}
Son D~T and Surowka P 2009 {\em Phys. Rev. Lett.\/} {\bf 103} 191601

\bibitem{neiman2011relativistic}
Neiman Y and Oz Y 2011 {\em Journal of High Energy Physics\/} {\bf 2011} 1--12

\bibitem{isachenkov2011chiral}
Isachenkov M and Sadofyev A 2011 {\em Physics Letters B\/} {\bf 697} 404--406

\bibitem{nair2012fluids}
Nair V~P, Ray R and Roy S 2012 {\em Physical Review D\/} {\bf 86} 025012

\bibitem{stephanov2012chiral}
Stephanov M and Yin Y 2012 {\em Physical review letters\/} {\bf 109} 162001

\bibitem{son2012berry}
Son D~T and Yamamoto N 2012 {\em Physical review letters\/} {\bf 109} 181602

\bibitem{banerjee2012constraints}
Banerjee N, Bhattacharya J, Bhattacharyya S, Jain S, Minwalla S and Sharma T
  2012 {\em Journal of High Energy Physics\/} {\bf 2012} 1--57

\bibitem{jensen2013thermodynamics}
Jensen K, Loganayagam R and Yarom A 2013 {\em Journal of High Energy Physics\/}
  {\bf 2013} 1--35

\bibitem{haehl2014effective}
Haehl F~M, Loganayagam R and Rangamani M 2014 {\em JHEP\/} {\bf 2014} 1--62

\bibitem{dubovsky2014effective}
Dubovsky S, Hui L and Nicolis A 2014 {\em Physical Review D\/} {\bf 89} 045016

\bibitem{monteiro2015hydrodynamics}
Monteiro G~M, Abanov A~G and Nair V 2015 {\em Phys. Rev. D\/} {\bf 91} 125033

\bibitem{mitra2022divergence}
Mitra A~K and Ghosh S 2022 {\em Physical Review D\/} {\bf 106} L041702

\bibitem{carter1979perfect}
Carter B 1979 Perfect fluid and magnetic field conservation laws in the theory
  of black hole accretion rings. {\em Active galactic nuclei\/} ed Hazard C and
  Mitton S (Cambridge: Cambridge University Press) pp 273--300

\bibitem{arouca2022quantum}
Arouca R, Cappelli A and Hansson T 2022 {\em arXiv preprint arXiv:2204.02158\/}

\bibitem{carter1988standard}
Carter B and Gaffet B 1988 {\em Journal of Fluid Mechanics\/} {\bf 186} 1--24

\bibitem{markakis2017conservation}
Markakis C, Ury{\=u} K, Gourgoulhon E, Nicolas J~P, Andersson N, Pouri A and
  Witzany V 2017 {\em Physical Review D\/} {\bf 96} 064019

\bibitem{seliger1968variational}
Seliger R~L and Whitham G~B 1968 {\em Proc. R. Soc. Lond. A\/} {\bf 305} 1--25

\bibitem{schutz1970perfect}
Schutz~Jr B~F 1970 {\em Physical Review D\/} {\bf 2} 2762

\bibitem{volovik1979poisson}
Volovik G and Dotsenko~Jr V 1979 {\em JETP Lett.\/} {\bf 29} 576--579

\bibitem{rieutord2006introduction}
Rieutord M, Dubrulle B and Gourgoulhon E 2006 {\em EAS Publ. Ser.\/} {\bf 21}
  43--79

\bibitem{khesin1989invariants}
Khesin B and Chekanov Y~V 1989 {\em Physica D: Nonlinear Phenomena\/} {\bf 40}
  119--131

\bibitem{bekenstein1987helicity}
Bekenstein J~D 1987 {\em The Astrophysical Journal\/} {\bf 319} 207--214

\bibitem{arnold2008topological}
Arnold V~I and Khesin B~A 2008 {\em Topological methods in hydrodynamics\/} vol
  125 (Springer)

\bibitem{webb2014local}
Webb G, Dasgupta B, McKenzie J, Hu Q and Zank G 2014 {\em Journal of Physics A:
  Mathematical and Theoretical\/} {\bf 47} 095502

\end{thebibliography}

%%%%%%%%%%%%%%%%%%%%%%
%%%%%%%%%%%%%%%%%%%%%%
%%%%%%%%%%%%%%%%%%%%%%

\end{document}